\documentclass{sf2a-conf2015}
\usepackage{graphicx}
\usepackage{hyperref}
\usepackage[]{natbib}  
\usepackage{epstopdf}

\def\BibTeX{{\rm B\kern-.05em{\sc i\kern-.025em b}\kern-.08em
    T\kern-.1667em\lower.7ex\hbox{E}\kern-.125emX}}
\bibpunct{(}{)}{;}{a}{}{,}  %%%%%%%%%%%%%  A&A bibliography style
\begin{document}

\TitreGlobal{SF2A 2015}

%%-----------------------------------------------------------------
%%      the top matter
%%

\title{Exposure-based Algorithm for Removing Systematics out of the CoRoT Light Curves}

\runningtitle{Systematics CoRoT}

\author{P. Guterman$^{1,}$}\address{Division Technique INSU, BP 330, 83507 La Seyne cedex, France}\address{ LAM (Laboratoire d'Astrophysique de Marseille), UMR 7326, F-13388 Marseille, France}

\author{T. Mazeh}\address{School of Physics and Astronomy, Tel Aviv University, Tel Aviv 69978, Israel}

%% IF Author3 has the same affiliation than Author1:
%\author{C.\,E. Author3$^1$}

%% IF Author3 has its own affiliation:
%\author{C.\,E. Author3}\address{Dept. of Chess, University of Games, 35101 Las Vegas, Monaco} 

%% IF Author3 has two affiliations, the one of Author1 and a second one:
\author{S. Faigler$^{3}$}

%% Keep this line, even if the page will be settled afterwards.
\setcounter{page}{1}

%%-----------------------------------------------------------------

\maketitle

%%-----------------------------------------------------------------
%%        The abstract
%% 
%%  Warning!  within the abstract:
%%  - do not use macros. 
%%  - do not use commands like: \cite, \citet, \citep ... etc.

\begin{abstract}
The CoRoT space mission was operating for almost 6 years, producing thousands of continuous photometric light curves. The temporal series of exposures are processed by the production pipeline, correcting the data for known instrumental effects. But even after these model-based corrections, some collective trends are still visible in the light curves.  We propose here a simple exposure-based algorithm to remove instrumental effects. The effect of each exposure is a function of only two instrumental stellar parameters, position on the CCD and photometric aperture. The effect is not a function of the stellar flux, and therefore much more robust. As an example, we show that the $\sim2\%$ long-term variation of the early run \textit{LRc01} is nicely detrended on average. This systematics removal process is part of the CoRoT \textit{legacy} data pipeline. 
\end{abstract}

%% Insert the keywords (to appear in the ADS indexing)
%% Keywords must be separated by a comma
\begin{keywords}
techniques: photometric, methods: data analysis 
\end{keywords}

%%-----------------------------------------------------------------

\section{Introduction}
\begin{figure}[!h]
  	\centering
  	\includegraphics[width=0.42\textwidth]{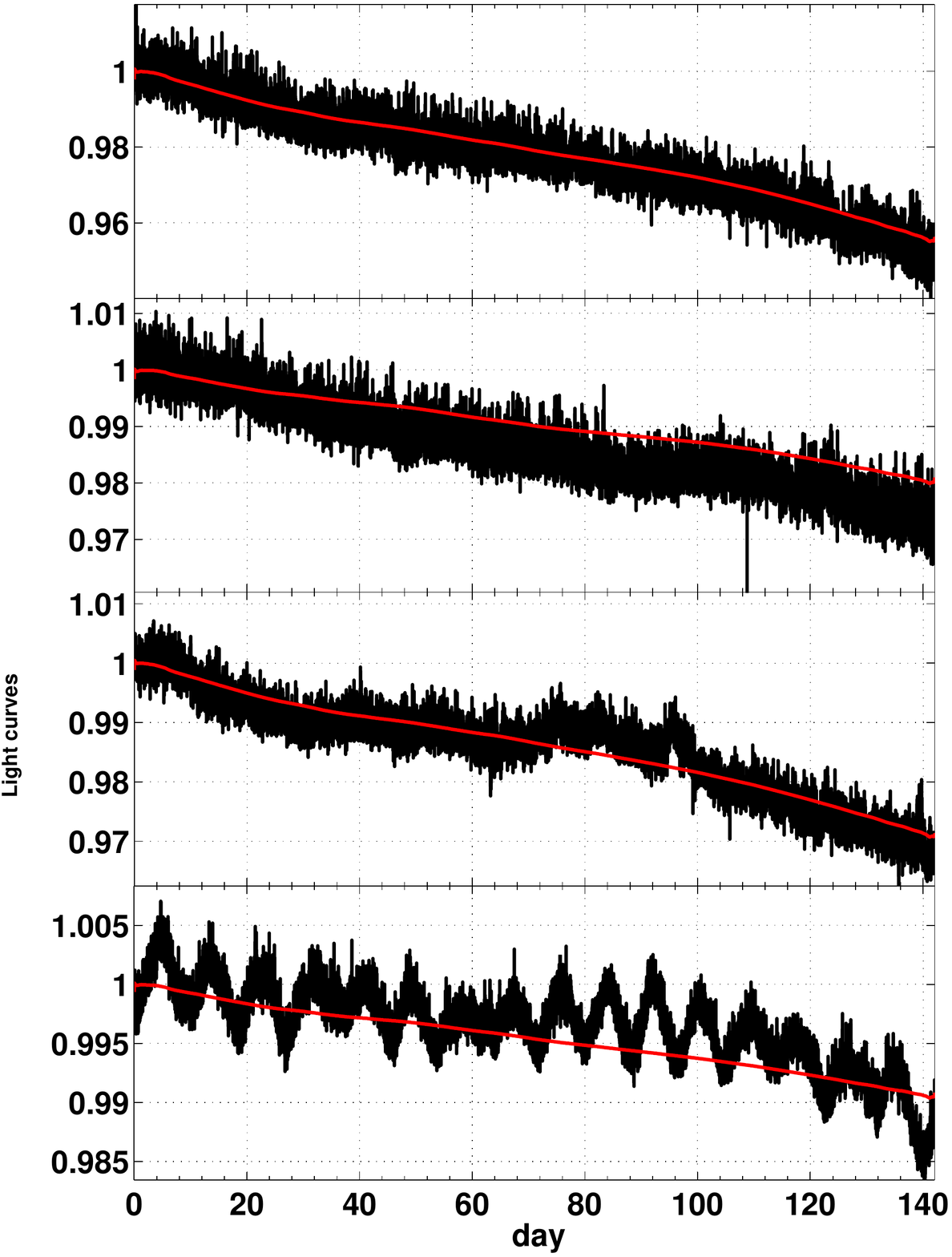}%      
  	\includegraphics[width=0.42\textwidth]{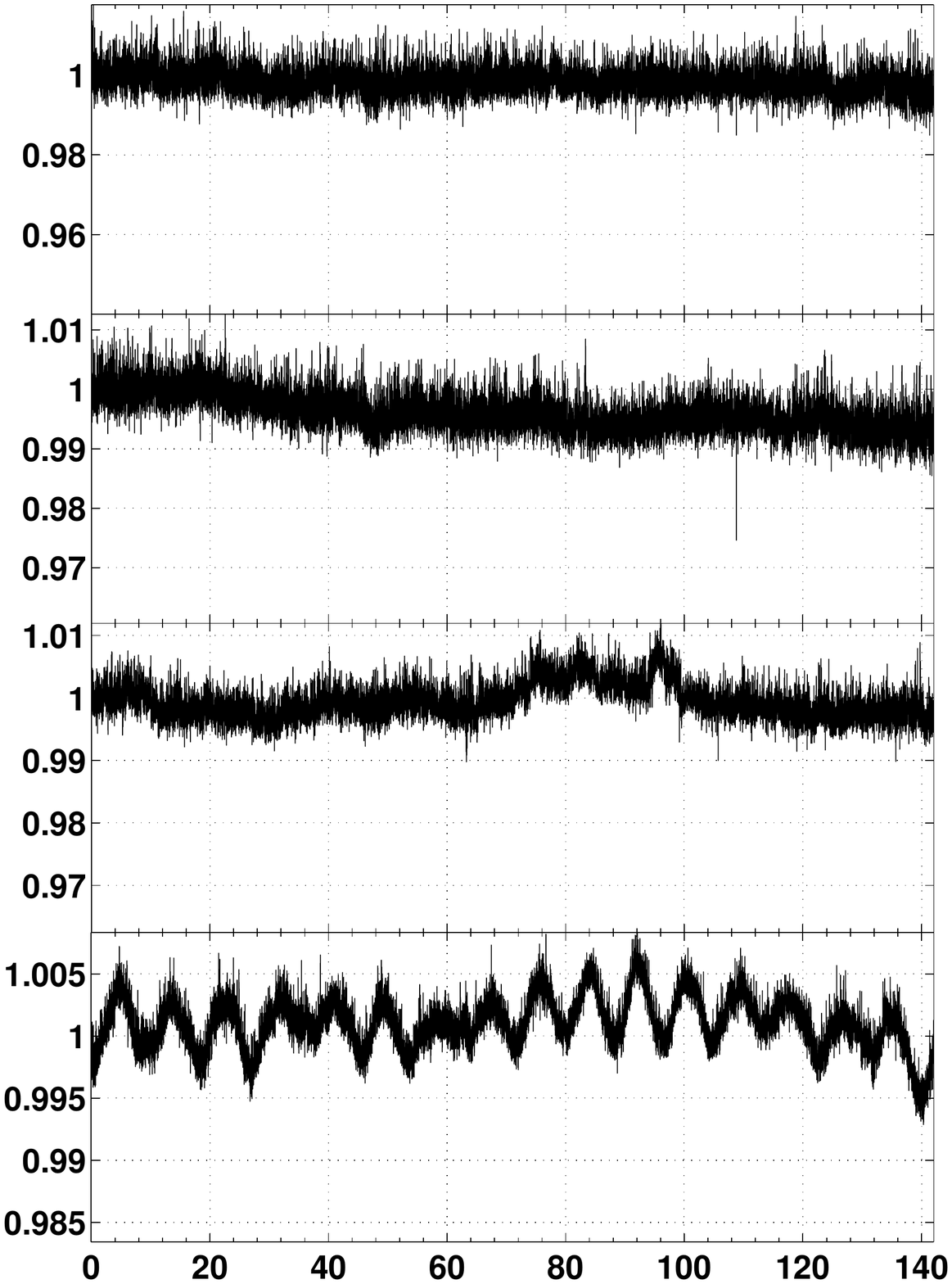}      
  	%% Note the ABSENCE of the extension .pdf  !
  	\caption{Corrected stellar trends. Left: Light curves without correction. The considered light curves are from run \textit{LRc01}, CCD 2. The black line is the normalized flux, the red line is the predicted systematic (see below). Right: same curves after correction.}
  	\label{guterman:fig1}
  \end{figure}

%%---------------------
  The CoRoT space mission \citep{Baglin2006} was operating for almost 6 years, producing thousands of continuous photometric light curves. The readout of each CCD exposure transfers simultaneously the flux of 6,000 stars. The temporal series of exposures are processed by the production pipeline, correcting the data for known instrumental effects, such as gain, background, jitter, EMI, SAA discarding, time corrections \citep{Samadi2006, Auvergne2009}. But even after these model-based corrections, some collective trends are still visible in the light curves (Fig.~\ref{guterman:fig1}). The flux gradually decreases with unknown shape and a different slope for each star. Previous work to correct these effects has been suggested, including MagZeP \citep{Mazeh2009}, that uses a zero-point magnitude correction, associated with the SysRem systematics algorithm \citep{Tamuz2005}. 
  \\
  
  %%
  %% Example of two figures side by side
  %%

  Algorithms for removing systematics consist of two parts: 1) identify the {\it effects} among a set of stars by combining all light curves like SysRem \citep{Tamuz2005}, see also \citep{Ofir2010}, finding  combination of a few representative stars \citep{Kovacs2005} or fitting a model for each exposure based on observational \citep{Kruszewski2003} or instrumental quantities \citep{Mazeh2009}, then 2) remove them by properly adapting them to each light curve. An effect is a pattern that appears among a large set of independent stars. Effects can be additive, multiplicative or follow any law that needs to be determined. In the common techniques, the effects are derived from a training set of stars using correlation methods like the iterative SysRem \citep{Tamuz2005}. The training set can be a properly selected subset of stars or even the whole set itself.
  \\
    
  After their global determination, the effects need to be scaled and subtracted from each of the light curves. Classical fitting techniques like least square are not satisfactory because the resulting coefficient is partly pulled by the light curve's natural shape and disturbs the scientific signal. For example, the gradual loss of sensitivity visible in Fig.~\ref{guterman:fig1} correlates with any long-period stellar variability, resulting in removing some real signal. To avoid this critical drawback we propose here a technique similar to MagZeP \citep{Mazeh2009} that fits the instrumental effects to each exposure independently. The effect of each exposure is a function of only two instrumental stellar parameters, position on the CCD and photometric aperture. The advantage is that the effect is not a function of the stellar flux, and therefore much more robust.
  \\
  
  This paper is structured as follows: Section \ref{sectMethod} describes the systematics removal method and its application to CoRoT, Section \ref{sectResult} reviews the derived effects and the performances of the method and Section \ref{sectSummary} summarizes and concludes this work.
  
\section{Method}
\label{sectMethod}
\begin{figure}[h]
	\centering
	\includegraphics[width=0.4\textwidth]{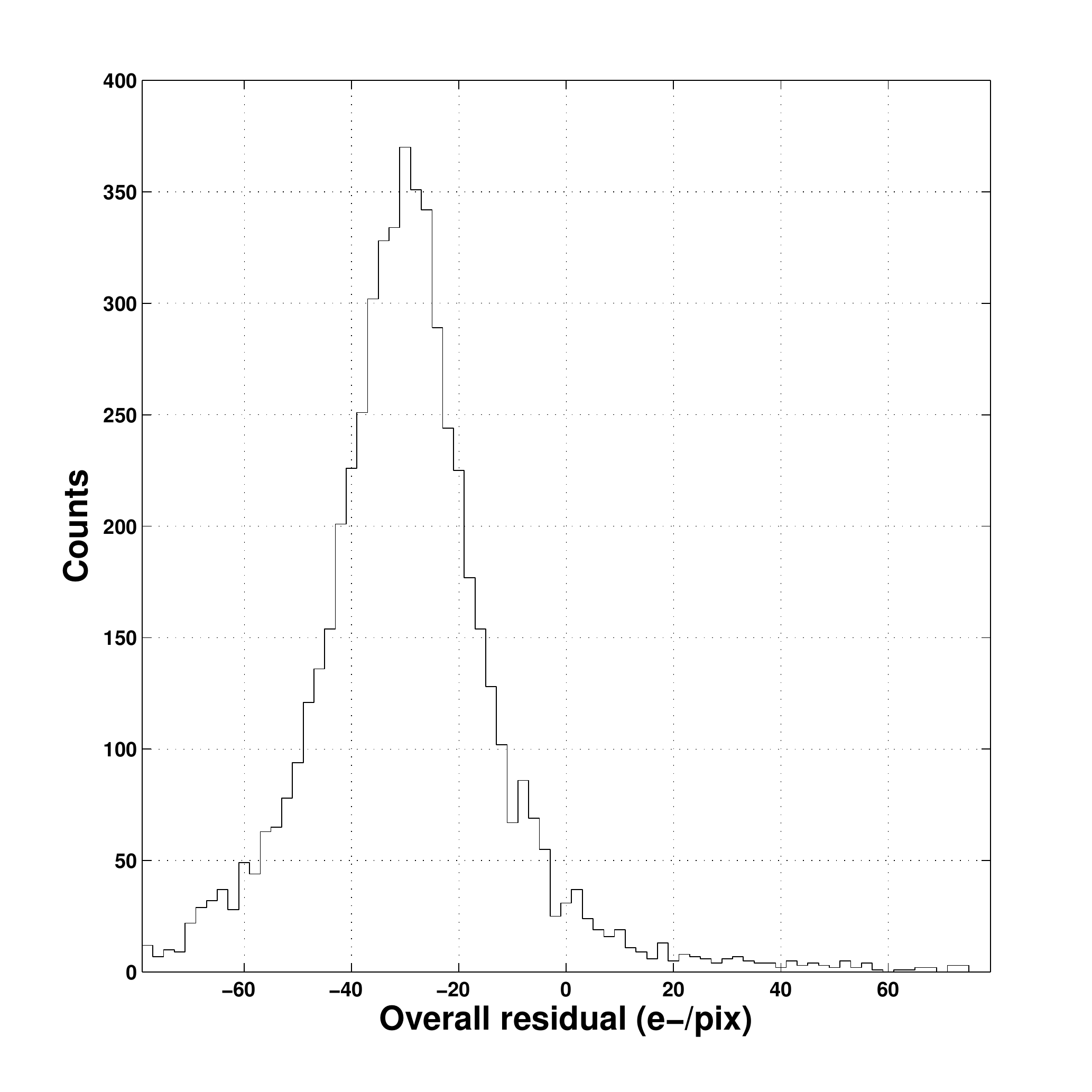}      	
	\includegraphics[width=0.4\textwidth]{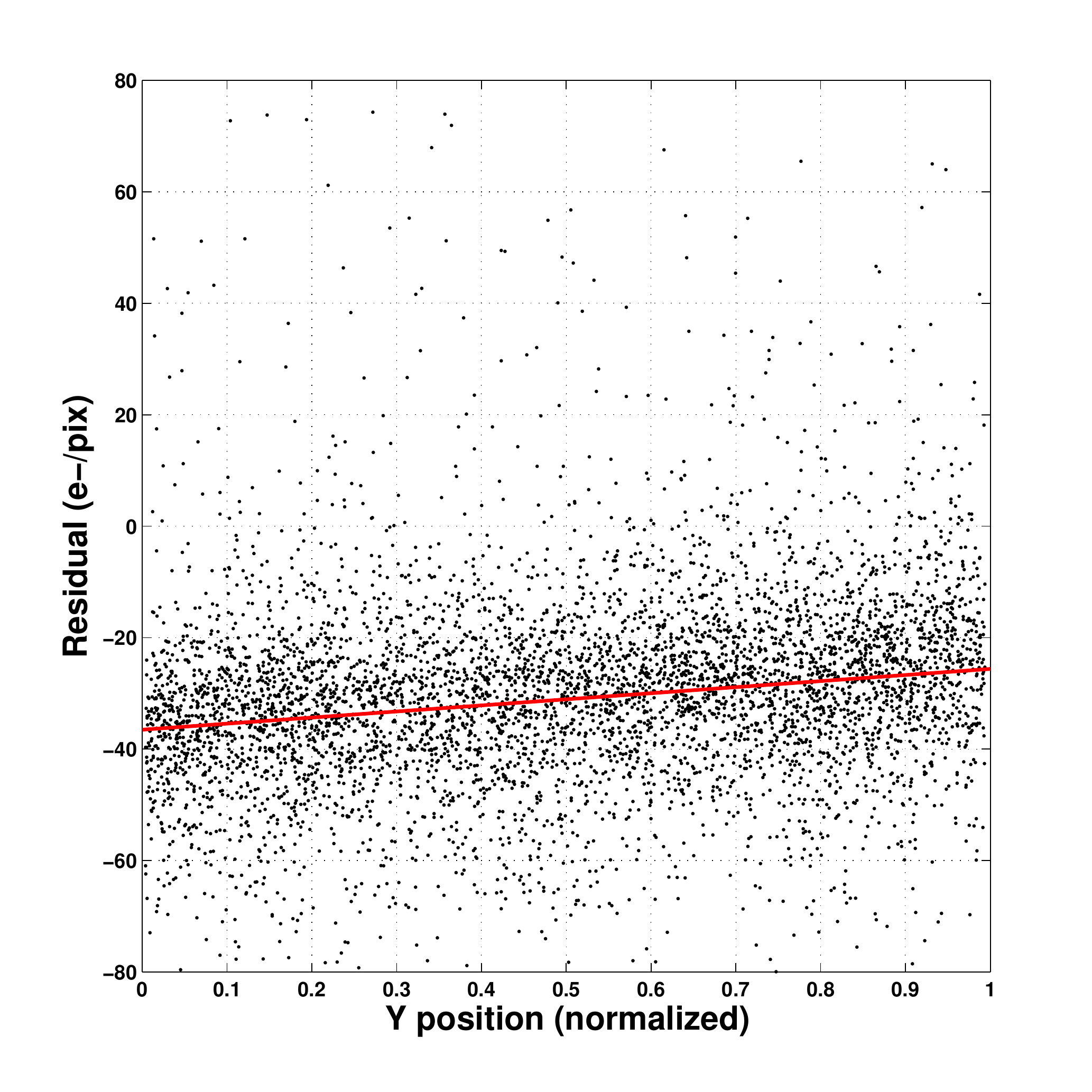}      	
	%% Note the ABSENCE of the extension .pdf  !
  	\caption{Systematic offset. Left: an histogram of all light curves residuals for an single exposure at the end of \textit{LRc01}. Right: same residuals as a function of $x$ positions. The red line is the linear approximation of the dependence.}
	\label{guterman:fig2}
\end{figure}

%%-------------------------
We compute the {\it residual} per pixel of each star by removing its zero point 
\begin{equation}
	r(t)=\frac{f(t)-\bar{f}}{m},
\label{eqresid}
\end{equation}
where $f$ is the star's flux and $m$ is the mask surface in pixels. The fluxes are divided by the photometric aperture size $m$ because systematics are at the pixel scale. The average flux $\bar{f}$ is integrated over the first 4 days of the run, before any drift could occur. Thus, all residuals are distributed around zero at the beginning of the run. Fig.~\ref{guterman:fig2} (left), which depicts the histogram of the stellar flux during the 4 last days of the run, shows that by the end of the run, the residuals are significantly below zero. The fluxes loose about 30$\mathrm{e^{-}/pix}$ on average after 140 days. This offset does not seem proportional to the flux itself, but rather linked to the position of the star on the CCD (right). Moreover, the dependence is close to linear, hence suggesting a model
\begin{equation}
	S_{ij} =C_i +A_ix_j+B_iy_j,
	\label{eqmodel}
\end{equation}
with $S_{ij}$ being the systematic offset of the $i^\mathrm{th}$ exposure of the $j^\mathrm{th}$ star located at position $x_j,y_j$ on the CCD. The coefficients $A_i$ and $B_i$ form the position dependence and $C_i$ is the common offset at exposure $i$. For each exposure we fit the three parameters $A_i$, $B_i$ and $C_i$ using a robust estimator. Next, we  smooth the derived $A$, $B$ and $C$ temporal curves (Fig.~\ref{guterman:fig3}), and then subtract the resulting $S_{ij}$ model from each exposure of each star.

\subsection{Processing}
Several steps are necessary to process the CoRoT data. 
\begin{enumerate}
	\item Resynchronising the run data 
	
	The CoRoT data is a collection of files, each containing the run light curve of a single star. We pack all the files of a run into a single matrix of $\mathrm{star}\times\mathrm{exposure}$ that contains the whole data of that run. The difficulty is that although simultaneously acquired, the time label of an exposure differs across files, depending on star position and roundoff errors among others. Consequently we had to gather all measurements within a common 4 $\sec$ interval across the whole file set as belonging to the same exposure.
	
	\item Binning to 512 $\sec$
	
	Some of the stars are sampled at a 32 $\sec$ cadence while the rest are at 512 $\sec$ cadence. A 512 $\sec$ measurement is the onboard concatenation of 16 successive 32 $\sec$ exposures. The timestamp is the center of the exposure interval in both cases. The present step consists on binning the data matrix to a common 512 $\sec$ time frame in the same way that CoRoT would have done onboard. 
	   
	\item Deriving the effects coefficients
	
	We compute the residual of each star (Eq.~\ref{eqresid}) and produce the residual matrix of the run. Then, for each exposure $i$ of that matrix, we estimate the coefficients $A_i,B_i$ and $C_i$ (Eq.~\ref{eqmodel}) using  robust multi-linear regression \citep{holland77}. This way, the procedure is insensitive to outliers due to spurious cosmic rays or stellar variability. Another benefit of such an estimator is that there is no need to select a training set.
	
	The resulting $A, B$ and $C$ temporal coefficients are then strongly smoothed using a 30 days sliding average to remove high-frequency noise caused by the fitting process and modeling imperfections. Fig.~\ref{guterman:fig3} shows the temporal evolution of the $A,B,C$ coefficients for \textit{LRc01}. 
	
\begin{figure}[h]
	\centering
	\includegraphics[width=0.7\textwidth]{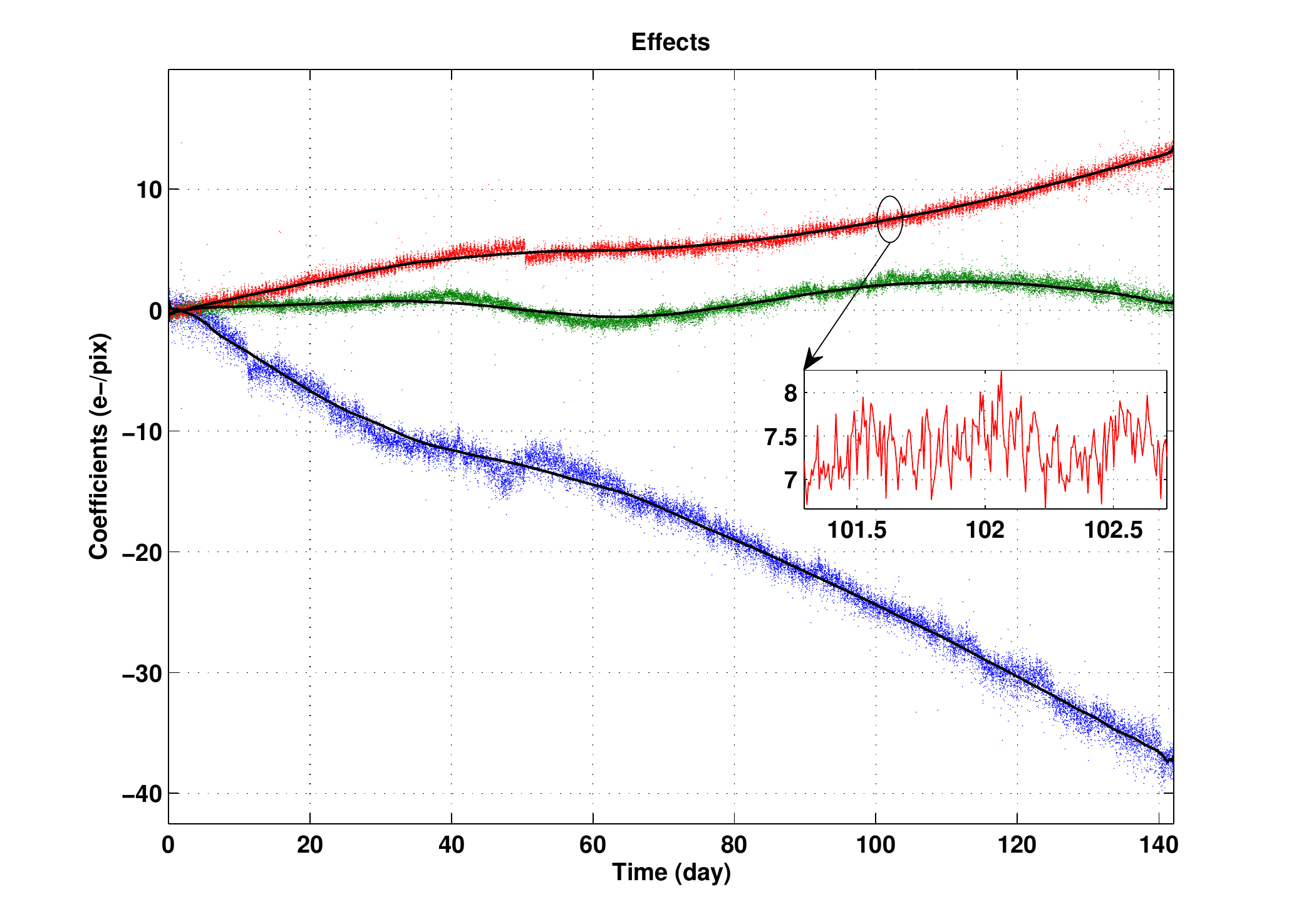}      	
	%% Note the ABSENCE of the extension .pdf  !
	\caption{Effects as a function of time for the long run \textit{LRc01}. Blue-offset $C$; green-$A$ ($x$ dependence); red-$B$ ($y$ dependence). The magnified section illustrates finer details on a shorter time scale which we ignore. The time is the number of days since the beginning of the run.}
	\label{guterman:fig3}
\end{figure}
		
	\item Removing the systematic effects model
	
	We derive the $S_{ij}$ matrix (Eq.~\ref{eqmodel}) and subtract it from the residual matrix. The resulting residuals are then reverted back to light curves through the inverse of Eq.~\ref{eqresid}. This part is performed by the production pipeline that stores the result into a specific extend of the legacy \textit{fits} files. This process takes place after the gap filling and the jumps corrections stages. 		
\end{enumerate}

%%
%% Example of single figure
%%

\section{Results}
\label{sectResult}
\subsection{Effects}
Fig.~\ref{guterman:fig3} shows the evolution of the model coefficients during the 142 days of run \textit{LRc01}. In blue, the common offset coefficient $C$ shows that all light curves gradually looses up to 37$\mathrm{e^{-}/pix}$ during the 142 days of the run. This long-term trend may reflect the loss of efficiency of the CCD, attributed to aging effects. In red (green), the $x$ ($y$) coefficients $A$ ($B$) show patterns of lower amplitude, probably caused by the star shift inside its mask, probably due to small rotational depointing or aberration. The larger value of the $B$ coefficient (red) relative to the $A$ coefficient (green) could come from stretching of the CoRoT PSF along the $x$ direction \citep{Llebaria2004}. Thus, a small displacement in the $y$ direction influences the signal proportionally more than the same displacement along the $x$ axis. 

A change of the CCD temperature is visible as a discontinuity at $t\sim50$ days, particularly in the red curve. Smaller details are visible down to the order of 1$\mathrm{e^{-}/pix}$ in the 1.5-day magnified section. The faster oscillations are the residual of the CoRoT satellite orbital period, namely 13.97 day$^{-1}$. A daily pattern variation is also clearly visible. Although interesting for analysis purposes, such details are removed from the operational coefficients by the smoothing process, because the model is not accurate enough for these patterns. 

\subsection{Performance}
\begin{figure}[h]
	\centering
	\includegraphics[width=0.65\textwidth]{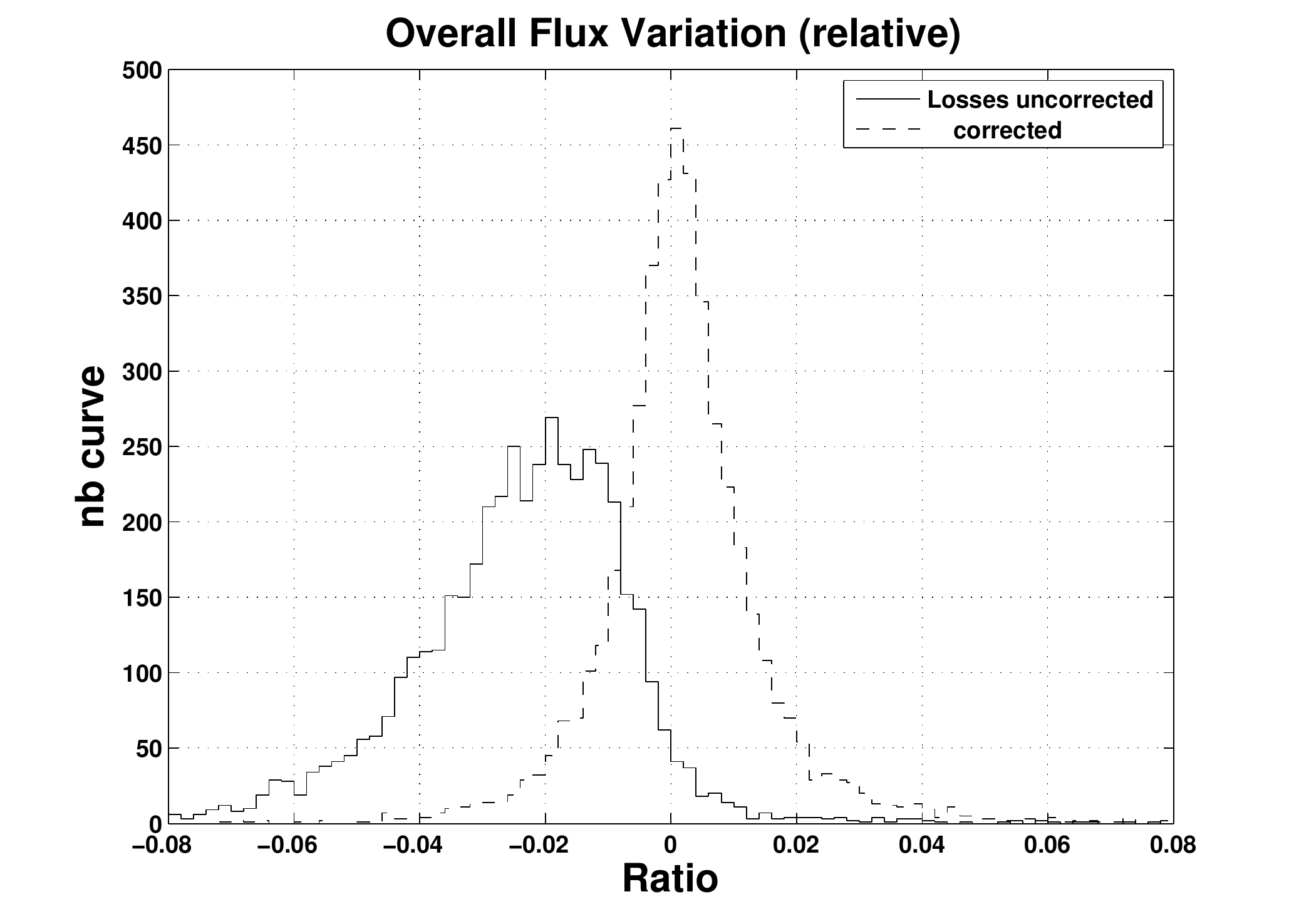}      	
	%% Note the ABSENCE of the extension .pdf  !
  	\caption{Results. Black line: histogram of flux losses at the end of run \textit{LRc01}. The loss is intended relatively to the initial flux. Dashed: same after correction.}
	\label{guterman:fig4}
\end{figure}

Fig.~\ref{guterman:fig4} shows the flux loss histograms in \textit{LRc01} before and after the correction, and illustrates that our method efficiently corrects the flux decrease. Before correction (solid line), the overall difference $\Delta=f_{\mathrm{end}}-f_{\mathrm{begin}}$ spreads around 2\% loss in 142 days. This 2\% loss is equivalent to $\sim1000\mathrm{e^{-}/pix}$, assuming an average mask area. After correction (dashed line), in addition to removal of the bias, the histogram is sharper. This reduction of differences between stars illustrates the effectiveness of the position based approach.

While this method is efficient on average, visual inspection of many stars reveals its limitations. For many stars, the systematics model nicely fits the light curve. However, for other stars over/under corrections can be seen. We tested several possibilities for explaining this. We checked the influence of the photometric masks geometry with collective depointing. For this, we used the full pixel images recorded before the run. We also checked the smearing due to readout pattern across columns. We even performed a blind search for correlations with combinations of available parameters like spectral type, magnitude, mask surface and others. Eventually we were not able to identify any additional factor that could explain the suspected over/under correction.  

\section{Summary and conclusions}
\label{sectSummary}
We present a simple method to remove systematics from the light curves of the CoRoT satellite without altering the scientific information. We apply a 3 parameter linear model per exposure to identify and correct most long-term systematics. The robust estimation algorithm allows to use the full information of the CoRoT sample of a run without selecting a training subset. The derived systematics only depend on the stellar position and mask area and not on the corresponding light curve. Consequently, no fitting based on the light curve itself can modify the real stellar variation, even when it resembles the systematics profile.

As an example, we show that the $\sim2\%$ long-term variation of the early \textit{LRc01} is nicely detrended on average, and the spread of stars variations is reduced. This systematics removal process is part of the CoRoT \textit{legacy} data pipeline. 

\section*{Acknowledgements}
This research has received funding from the European Community's
Seventh Framework Programme (FP7/2007-2013) under grant-agreement
numbers 291352 (ERC) 

\bibliographystyle{aa}  % A&A bibliography style file (aa.bst)
%%\bibliography{sf2a-template} % your references in file: Yourfile.bib

% 2006ESASP1306..317S Extraction of the Photometric Information: Corrections, Samadi, R.; Fialho, F & Al

%
\end{document}